# High-resolution long-working-distance reference-free holographic microscopy exploiting speckle-correlation scattering matrix

YoonSeok Baek[1,2], KyeoReh Lee[1,2] and YongKeun Park[1,2,3*]

[1]Department of Physics, Korea Advanced Institute of Science and Technology (KAIST), Daejeon 34141, Republic of Korea.

[2]KAIST Institute for Health Science and Technology, KAIST, Daejeon 34141, Republic of Korea.

[3]Tomocube, Inc., 48, Yuseong-daero 1184beon-gil, Yuseong-Gu, Daejeon 34051, Republic of Korea,

*yk.park@kaist.ac.kr

**Abstract**

Using conventional refraction-based optical lens, it is challenging to achieve both high-resolution imaging and long-working-distance condition. To increase the numerical aperture of a lens, the working distance should be compensated, and *vice versa*. Here we propose and demonstrate a new concept in optical microscopy that can achieve both high-resolution imaging and long-working-distance conditions by utilising a scattering layer instead of refractive optics. When light passes through a scattering layer, it creates a unique interference pattern. To retrieve the complex amplitude image from the interference pattern without introducing a reference beam, we utilised a speckle-correlation scattering matrix method. This property enables holographic microscopy without any lens or external reference beam. Importantly, the proposed method allows high-resolution imaging with a long working distance beyond what a conventional objective lens can achieve. As an experimental verification, we imaged various microscopic samples and compared their performance with off-axis digital holographic microscopy.

**Introduction**

In optical microscopy, the performance of an objective lens is greatly affected by the size of an aperture. A finite lens aperture forces a fixed relationship between a numerical aperture (NA) and a working distance. Thus, when the aperture size is fixed, there is a trade-off between NA improvement and improvement in working distance, which becomes a critical issue when high-resolution and long-working distance cannot be compromised, such as imaging of a thick tissue, an extracellular matrix[1], or a microfluidic channel[2].

A straightforward solution is to expand the physical size of the aperture. In reality, however, this is impractical, because aberration becomes severe with increase in the aperture size, and an objective lens is composed of a series of different lenses to compensate aberration. These make increasing the aperture size challenging and costly, and several alternative approaches have been introduced. For example, Fourier ptychography[3,4] can mitigate the issue by simulating a large aperture imaging system. It allows high-resolution imaging with a low-NA objective lens that typically has a long working distance. However, it requires a thin sample[5], and a heavy computational workload is unavoidable. Synthetic aperture microscopy enables the reconstruction of a high-NA image from multiple low-NA images gathered at various illumination angles[6]. However, this technique is also limited for use in thin samples and requires multiple measurements[7].

A possible alternative to the existing solution is to exploit multiple scattering of light in a disordered medium. Multiple scattering is a deterministic process which is described by a scattering matrix or a transmission matrix (TM) in a transmission geometry[8, 9, 10]. Recent studies have demonstrated that a scattering layer can be used for focusing and imaging once the TM is calibrated[11, 12, 13, 14, 15]. TM-based imaging has been adopted in microscopy, yielding an improved resolution beyond what a refraction-based imaging system with an objective lens can achieve[16, 17]. However, TM-based imaging methods require a complicated interferometry system. Alternatively, microscopic imaging through a scattering layer has been demonstrated[18] by exploiting intensity correlation techniques[19, 20]. However, its use is limited to imaging the intensity of an object at a fixed plane.

Here we demonstrate a high-resolution long-working-distance reference-free holographic microscopy technique using a scattering layer. The scattering layer replaces the role of an objective lens and that of an interferometer simultaneously (Fig.1). Exploiting the speckle-correlation scattering matrix (SSM) technique[21], the scattering layer can be utilised as a high-NA imaging lens which creates long working distance conditions. By exploiting the randomness of a multiple scattering layer, SSM uniquely determines an incident field from the scattered intensity image of after passing through a scattering layer. This allows holographic imaging of a microscopic object without introducing a reference beam for off-axis interferometry. We experimentally demonstrate a high-NA (0.7) with a working distance of 13 mm, which significantly exceed the capabilities of existing commercially available high-NA and long-working-distance objective lenses. The performance of the proposed method is demonstrated by imaging various microscopic samples.

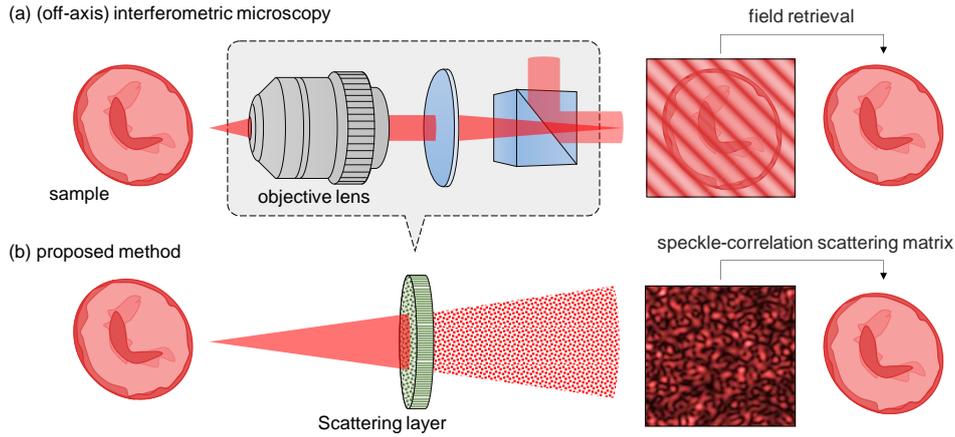

**Figure 1: Comparison between conventional interferometric microscopy and the proposed method.** (a) Schematic of off-axis interferometric microscopy. An objective lens is used to collect diffracted light from a sample with high numerical aperture. In addition, an external reference beam should be introduced to perform interferometry. The complex amplitude at the sample plane is retrieved from the measured interferogram. (b) A schematic of the proposed method is shown. In the proposed method, the optical components are replaced with a single scattering layer. The scattering layer generates a unique interference pattern, from which the complex amplitude is retrieved using the speckle-correlation scattering matrix technique.

**Results**

**Principle**

The principle of the proposed method is based on SSM. SSM is a method to retrieve the complex amplitude of an incident light beam from an intensity image of light transmitted through a scattering layer. SSM essentially utilises the TM information of a scattering layer. The TM, denoted as $T$, describes the relationship between an incident field $E_i$ and a transmitted field $E_t$ as $E_t = TE_i$. In order to determine an incident field, one must know the transmitted field, or vice versa. However, SSM provides a method to retrieve the incident field directly from the intensity image of the transmitted field[21]. The random distribution of amplitude and phase information in a speckle field enables a mathematical relation between the fourth-order and second-order moments of the optical field, also known as the Isserlis' theorem[22, 23, 24, 25].

For an incident field $x = \sum_n^N \alpha_n k_n$, where $k$ is a spatial frequency or generally an arbitrary orthonormal basis, the corresponding field after a scattering layer can be deterministically described using the TM of the scattering layer as $y = Tx = \sum_n^N \alpha_n t_n = \sum_n^N \alpha_n T k_n$. To apply SSM, a $Z$ matrix can be defined as

$$Z_{nm} = \frac{1}{\langle |t_n|^2 \rangle_r \langle |t_m|^2 \rangle_r} \left( \langle t_n^* t_m y^* y \rangle_r - \langle t_n^* t_m \rangle_r \langle y^* y \rangle_r \right), \tag{1}$$

where $y^*y$ corresponds to an intensity image of the transmitted field and $\langle \ \rangle_r$ indicates a spatial average. Note that the Z matrix contains the TM and the measured intensity image. For a strong scattering medium whose TM is described as a Gaussian random matrix[10, 26], the fourth-order moment of the Z matrix can be rewritten in terms of second order moments, according to the Isserlis' theorem. The fourth-order moment can be rewritten as

$$Z_{nm} = \alpha_n \alpha_m^* + \frac{\langle t_n^* y^* \rangle_r \langle t_m y \rangle_r}{\langle |t_n|^2 \rangle_r \langle |t_m|^2 \rangle_r}, \tag{2}$$

where $\alpha_n = \langle t_n^* y \rangle_r / \langle |t_n|^2 \rangle_r$. The second term in Eq. (2) is negligible as the columns of the Gaussian random TM are uncorrelated to each other, i.e., $\langle t_n t_m^* \rangle_r \sim \delta_{nm}$ and $\langle t_n t_m \rangle_r \sim 0$. In practice, however, the second term can be neglected with an additional condition, as the averaging over all space is impractical due to the limited number of pixels in a detector. This additional condition is to keep the number of detection modes (the number of pixels in a detector) larger than the number of input modes (or the number of basis states for the incident field)[21]. See Supplementary Note 1 for detailed information. If the above conditions are met, Eq. (2) is written simply as $Z = \alpha \alpha^*$, whose singular vector corresponds to the incident field.

In principle, there is a single non-zero singular value, and a corresponding singular vector is used to reconstruct the incident field. However, in the presence of noise, there is more than one non-zero singular value. In this case, the singular vector having the largest singular value is used instead.

**Experimental demonstration**

The experimental setup is shown in Fig. 2(a). A scattering layer is placed at the objective lens position. Detection is achieved with the scattering layer, a linear polarizer and a CCD. A digital micromirror device (DMD) and a condenser lens are used to calibrate the TM of the scattering layer (Fig. 2b & c).

To demonstrate the performance of the proposed method, we imaged various samples including a 1951 USAF resolution target (#59-153, Edmund optics, New Jersey, USA), a quantitative phase microscopy target (Benchmark technologies, Massachusetts, USA), and a human red blood cell (RBC). The experiments were conducted in the following order. First, the TM of a scattering layer was calibrated (see Methods). After the calibration process, samples were placed in the calibrated field of view (FOV). Then the calibrated FOV was illuminated with a plane wave (see Supplementary Note 2). The corresponding scattered light images that passed through the scattering layer were recorded with a CCD (see Methods). Finally, the complex amplitude at the sample plane was retrieved using SSM.

The experimental results are shown in Fig. 3. The leftmost column in Fig. 3 shows images of the scattered light after passing through the scattering layer for different samples. Each sample generates a completely different speckle pattern. First, we applied SSM and an iterative algorithm to the images to retrieve the complex amplitudes (see Methods). The results provide visualisations of samples, but speckle noise is present in both the amplitudes and phases (Fig. 3(b)). The speckle noise arises from the low SNR at weak speckles. The NA at this point is 0.6, which is determined by TM calibration. The results following aperture synthesis[6, 27, 28] are shown in Fig. 3(c), where speckle noise is greatly suppressed and clear images of the samples are shown. This leads to an effective NA of 0.7, with an increase of 0.1 (see Methods for details).

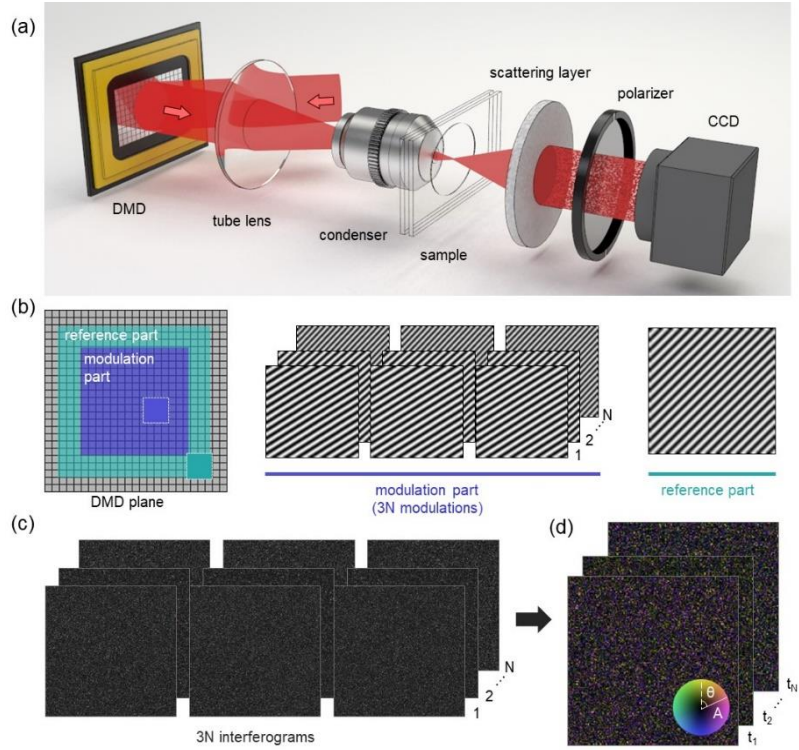

**Figure 2: Experimental setup and TM calibration.** (a) Experimental setup with a HeNe laser (633 nm) used as a light source. The arrows indicate the direction of light propagation. (b) DMD patterns for TM calibration. The central part of the DMD is modulated with $3N$ different patterns, where $N$ is the number of total input modes. In the peripheral part, a static pattern is projected, which serves as a reference field. (c) Measured speckle intensity patterns. Three speckle patterns per input mode were measured. (d) TM calculated from the measured speckle intensity patterns.

The first row of Fig.3 shows results for an amplitude mask from the 1951 USAF resolution target (the 3rd element of the 8th group). The results agree with the standard width of 1.55 μm. The second row corresponds to a quantitative phase resolution target, which has a height of 200 nm and a refractive index of 1.52. It induces a phase delay of 1.03 rad at 633 nm, which agrees with the measured phase value. The third row shows results from a human red blood cell. The red blood cell was diluted with DPBS and sandwiched with a cover slide. The measured phase image agrees well with the known shape of an RBC.

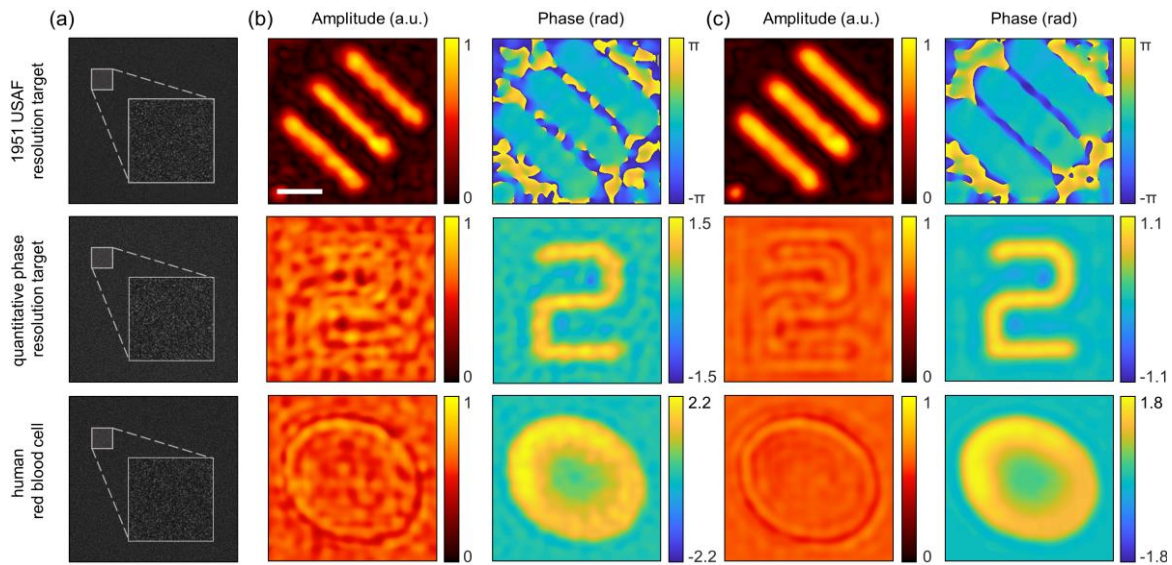

**Figure 3. Experimental results.** (a) Speckle intensity patterns generated using a scattering layer. (b) Complex amplitude images initially retrieved using SSM and the iterative algorithm applied to (a). (c) Final complex amplitudes retrieved after applying the synthetic aperture method to (b). The first row shows results from a 1951 USAF resolution target (the 3rd element in the 8th group). The second row shows result from the quantitative phase resolution target. The third row corresponds to a human red blood cell. The scale bar is 3 μm.

**High-resolution imaging at a long working distance**

The proposed method enables high-resolution imaging at a long working distance, as shown in Fig. 4(a). Since light hitting any position of the surface of the scattering layer undergoes multiple scattering, high spatial frequency information can be measured, even at a long working distance. As a demonstration, we imaged a 5 μm diameter silica bead (44054, Sigma-Aldrich, Missouri, USA) with the present method. Then we imaged the same sample with off-axis interferometric microscopy[29] using a long working distance objective (LUCPLFLN 60X, Olympus, Tokyo, Japan) (see Supplementary Figure 3 for the experimental setup of the off-axis interferometric microscope). The silica beads are immersed in water and sandwiched between two coverslips.

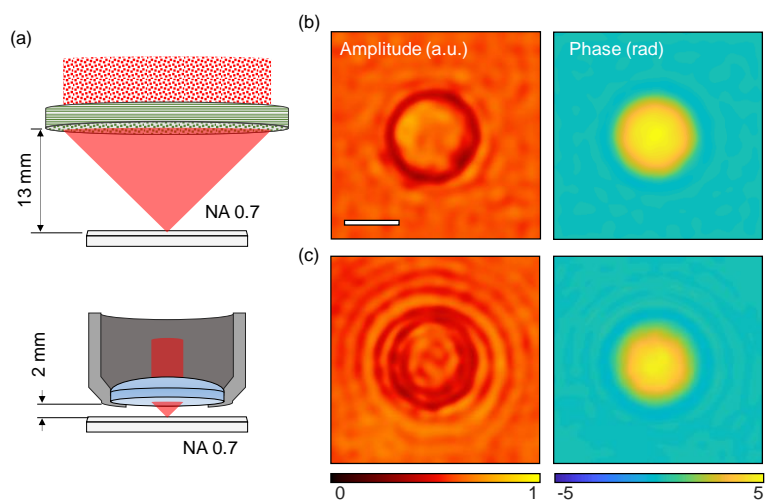

**Figure 4. Comparison of the proposed method and off-axis interferometric microscopy.** (a) The imaging conditions of the proposed method (above) and those in off-axis interferometric microscopy using a long-working-distance objective lens (below). In this experiment, the proposed method has an NA of 0.7 with a 13 mm working distance. The long working distance objective lens has an identical NA (0.7) and 2 mm working distance. (b) The measured complex amplitude for a 5 μm diameter silica bead using the proposed method. (c) Results from the same sample imaged using off-axis holographic microscopy with a long working distance objective lens. The scale bar is 3 μm.

The results from the proposed method and those from off-axis interferometric microscopy are shown in Fig. 4(b) and Fig. 4(c), respectively. The proposed method shows a clear image of a sample. On the contrary, off-axis interferometric microscopy suffers from objective lens aberration[30]. The aberration results in distortion in both the amplitude and phase images. In the demonstration, both methods have the same NA of 0.7. However, the proposed method provides a significantly long working distance of 13 mm. This is in contrast with the 2 mm working distance used with the commercial objective lens. The proposed method can provide a higher NA and a longer working distance by enlarging the size of the scattering layer. We note that the demonstrated high-resolution imaging with a long working distance condition is far beyond the capabilities of existing commercially available high-NA and long working distance objective lenses.

**Discussion**

In summary, we have demonstrated a microscopic technique that enables imaging with high numerical aperture and long working distance simultaneously. Exploiting the SSM approach, a high-resolution microscopic image of a sample is obtained from a single-shot measurement of a speckle intensity pattern, which is obtained in a simple optical setup composed of a scattering layer and an image sensor. The scattering layer serves both as an objective lens and an interferometer, and no other optical elements are required for measuring holographic images. With the proposed method, complex amplitudes from various samples are successfully measured from intensity images transmitted through the scattering layer. Experimental verification has been conducted with a USAF resolution target, a phase target, and an RBC.

This method provides a unique advantage in quantitative phase imaging and extends the applicability of a scattering layer as an optical element. For example, the proposed method provides highly flexible imaging conditions. The FOV, the NA, or the working distance can be modified through simple calibration. In addition, calibration can be performed as needed with the same setup used for imaging. Therefore, both calibration and imaging are accomplished without an external reference arm. Consequently, the proposed method enables imaging under different conditions without changing the experimental setup.

The proposed method fundamentally differs from holographic imaging methods using a scattering layer reported thus far. Previous methods focused on the use of a memory effect[31, 32, 33], the direct use of a TM, or the use of speckle intensity correlation[34]. The methods based on the memory effects [20, 35] require pseudo-thermal light sources or an illumination scan in addition to a phase-retrieval algorithm. In addition, they can retrieve 2D intensity images, not optical fields. For the conventional method based on the TM[16, 17] or intensity, the use of an interferometer with an external reference beam is inevitable, which makes an optical system significantly more complicated and limits practical applications. On the contrary, the proposed method only requires a scattering layer, and the incident field information is retrieved from an intensity image of the light passed through a scattering layer. Thus, the proposed method allows greater flexibility in the experimental setup.

One of the limitations of the present method is that it requires calibration of a scattering layer. However, due to the deterministic nature of multiple light scattering in complex media, the calibrated scattering layer has a long lifetime unless provided no changes occur in the scattering medium. Once calibrated, a scattering layer can be used as a high NA and long working distance lens. In laboratory conditions, the calibrated information for a scattering later has been used for 24 hours without additional calibration (see Supplementary Figure 4). For high-throughput generation of scattering layers, the photopolymerization of a calibrated scattering layer[36] or an engineered scattering layer[37] can also be used with this method.

It should be emphasised that the capability of the method is not restricted by the condition imposed on the number of input and detection modes. This condition appears to limit the amount measurable information as the number of detection modes cannot be increased indefinitely. However, one can work under this condition using various methods. For example, the NA of the proposed method can be extended beyond the calibrated NA using a synthetic aperture with a large illumination NA[16]. The FOV can also be extended with multiple TMs, each of which corresponds to different FOVs. The feasibility and performance of the method can be further improved with a specially designed scattering layer whose TM is calibrated in advance[37]. The optimal design can lead to little or no need of the post processing, including the iterative algorithm and synthetic aperture, by minimising the correlation between the TM columns.

**Methods**

**Experimental setup**

The experimental setup is shown in Fig. 2a. A collimated HeNe laser (HNL210L, Thorlabs, New Jersey, USA) illuminates a DMD (DLi4130, Digital Light Innovations, Texas, USA). The incident light is diffracted by a grating pattern displayed on the DMD. The diffracted light is captured by a plano-convex lens (f = 300 mm). Only the first order diffraction passes through the condenser lens (UPLFLN60X, Olympus, Tokyo, Japan). Other diffraction orders are filtered out by the back aperture of a condenser. As a result, only first order diffraction is imaged in a sample plane. This light illuminates a sample and propagates onto the scattering layer. The scattered light passes through a linear polarizer to generate a speckle pattern at the CCD (Lt365R, Lumenera, Ontario, Canada).

**Scattering layer fabrication**

A rutile paint was made by mixing rutile nanoparticles (637262, Sigma-Aldrich, Missouri, USA), resin (RSN0806, DOW CORNING, Michigan, USA) and solvent (toluene 99%) in a ratio of 0.5 g : 1 mL : 10 mL. The paint was sonicated for 10 min to disperse the rutile nanoparticles. Then the paint was applied on both sides of a 25 mm diameter round coverslip by using an air brush (DH-125, Sparmax, Taipei, Taiwan). Finally, the painted coverslip was baked (100°C, 10 min) to cure the resin. To ensure uniform speckle intensity, a ground glass diffuser (DG10-120, Thorlabs, New Jersey, USA) is used in addition to the rutile painted layer.

**TM calibration**

During TM calibration, the DMD is divided into two parts (Fig. 2(b)). One is a modulation part at the centre of the DMD. The modulation part corresponds to a FOV to be calibrated and it projects grating patterns with different spatial frequencies. As only the 1st order diffraction passes through the condenser lens, the grating patterns modulate the incident angle of light in the sample plane. The other is a reference part surrounding the central area. The reference part projects a stationary grating pattern throughout the calibration process, providing a static reference field. To maximise the visibility of the interferograms, the size of reference part is positioned such that it has the same surface area as the modulation part.

The total number of patterns used for calibration is determined by the number of input modes. For each input mode, three different patterns are used, resulting in phase shifts of 0, $2\pi/3$, $4\pi/3$. In addition, the intensity image of a reference field is imaged by turning off the modulation part. Thus, a total of $3N+1$ grating patterns are used. The number of input modes depends on the area of FOV, denoted as $A_{FOV}$, and NA to be calibrated. For example, in the experiment the number of input modes is expressed as $N \sim \pi A_{FOV} (NA/\lambda)^2$. From the $3N$ interferograms, we reconstructed the TM by a least-squares matrix inversion[38] as

$$\begin{pmatrix} |R|^2 + |t_n|^2 \\ R^* t_n \\ R t_n^* \end{pmatrix} = \begin{bmatrix} 1 & e^{i0} & e^{-i0} \\ 1 & e^{i2\pi/3} & e^{-i2\pi/3} \\ 1 & e^{i4\pi/3} & e^{-i4\pi/3} \end{bmatrix}^{-1} \begin{pmatrix} I_0 \\ I_{2\pi/3} \\ I_{4\pi/3} \end{pmatrix}. \qquad (3)$$

We consider the second element ($R^* t_n$) to be a calibrated TM. Note that the columns of the calibrated TM are multiplied with the complex conjugate of the reference field, making it different from the columns of the actual TM ($t_n$). Nevertheless, using $R^* t_n$ or $t_n$ yield identical Z matrices, although each will result in a different normalization factor[10] $|R|^2$. This normalization factor is the intensity image of the reference field measured during calibration. Therefore, we can obtain the Z matrix of the ideal TM from the calibrated TM and the intensity image of the reference field.

In the experiment, the total number of input modes used in the calibration is 657, which corresponds to a FOV of 15 × 15 μm and a NA of 0.6 for a HeNe laser (633 nm). The number of detection modes was 502,400. The TM calibration process took 394.2 seconds using 1971 4 bit patterns.

**Iterative algorithm**

The iteration algorithm is processed as follows. The singular vector ($\alpha$) of the Z matrix having the largest singular value is obtained by singular value decomposition. This singular vector is multiplied into the calibrated TM ($R^* t_n$). This gives a transmitted field ($R^* y$), whose intensity should be identical to the speckle image of a sample multiplied by the intensity image of the reference field ($|R|^2 |y|^2$). Therefore, we replace the amplitude of the transmitted field with the square root of the multiplied images. Finally, the replaced field is multiplied with the inverse TM. This, in turn, gives an updated singular vector. This process is repeated until no significant update is made to the singular vector. In the experiment, the iteration was set to stop when the correlation between an N-1-st singular vector and an N-th singular vector has a correlation above 0.99999. In the experiment, all data converged within 20 steps and required less than a minute of computation time.

### Synthetic aperture for noise reduction

First, the complex amplitudes at the sample plane under different illumination angles are measured using SSM and the iterative algorithm. Then we add the measured complex amplitudes together to gain a coherent summation of the signal and an incoherent summation of noise. Since the measured complex amplitudes are shifted differently in Fourier space, such shifts are compensated before the signals are added. For small changes in the illumination angle, the contribution of noise decreases as the number of illuminations (N) increases, i.e., $\sigma_{noise} \propto 1/\sqrt{N}$. In the experiment, 21 different illuminations were used with a maximum illumination NA of 0.1. This gives an increase in the NA by 0.1, resulting an effective NA of 0.7.

### Acknowledgement

This work was supported by KAIST, BK21+ program, Tomocube, and National Research Foundation of Korea (2015R1A3A2066550, 2017M3C1A3013923, 2014K1A3A1A09063027